\documentclass[
showpacs,
floatfix,
aps,prb,amsmath,
twocolumn,
superscriptaddress,
eqsecnum
]{revtex4}

\usepackage{epsfig}
\usepackage{bm}
\usepackage{epsf}
\usepackage{array}

\newcommand{\be}{\begin{equation}}
\newcommand{\ee}{\end{equation}}

\newcommand{\bea}{\begin{eqnarray}}
\newcommand{\eea}{\end{eqnarray}}

\newcommand{\bes}{\begin{split}}
\newcommand{\ees}{\end{split}}

\newcommand{\req}[1]{Eq.~(\ref{#1})}
\newcommand{\reqs}[1]{Eqs.~(\ref{#1})}
\newcommand{\rref}[1]{(\ref{#1})}
\renewcommand{\vec}[1]{{\bm #1}}
\newcommand{\Tr}[1]{\mathop{{\rm Tr}\left\{ #1\right\}}}
\newcommand{\tr}{\mathop{\rm Tr}}

\newcommand{\ep}{\varepsilon}

\renewcommand{\Im}{\mathop{\rm Im}}
\renewcommand{\Re}{\mathop{\rm Re}}

\newcommand{\spinmat}[1]{\hat{#1}}
\newcommand{\unitmatrix}{\hat{\bf 1}}
\newcommand{\magn}{h}
\newcommand{\df}{\mathcal{F}}

\newcommand{\mls}{\delta_1}

\begin{document}

\title{Stochastic dynamics of magnetization in
a ferromagnetic nanoparticle out of equilibrium}
\author{Denis M. Basko}
\affiliation{International School of Advanced Studies (SISSA), via Beirut 2-4, 34014 Trieste, Italy }
\author{Maxim G. Vavilov}
\affiliation{Department of Physics, University of Wisconsin, Madison, WI 53706, USA }

\date{September 15, 2008}

\begin{abstract}
We consider a small metallic particle (quantum dot) where
ferromagnetism arises as a consequence of Stoner instability. When
the particle is connected to electrodes, exchange of electrons
between the particle and the electrodes leads to a temperature- and
bias-driven Brownian motion of the direction of the particle
magnetization. Under certain conditions this Brownian motion is
described by the stochastic Landau-Lifshitz-Gilbert equation. As an
example of its application, we calculate the frequency-dependent
magnetic susceptibility of the particle in a constant external
magnetic field, which is relevant for ferromagnetic resonance
measurements.
\end{abstract}

\maketitle

\section{Introduction}

The description of
fluctuations of the magnetization in small ferromagnetic particles
pioneered by Brown\cite{Brown1963}  is based on
the Landau-Lifshitz-Gilbert (LLG) equation~\cite{LL35,Gilbert55}
with a phenomenologically added stochastic term. This approach has
been widely used: just a few recent applications are a
numerical study of the dynamic response of the magnetization to the
oscillatory magnetic field,\cite{Palacios1998} a numerical study of
ferromagnetic resonance spectra,\cite{Usadel2006} study of
resistance noise in spin valves,\cite{Foros2007} and a study of the
magnetization switching and relaxation in the presence of anisotropy
and a rotating magnetic field.\cite{Denisov2007}

In equilibrium the statistics of stochastic term in the
LLG equation can be simply written from the
fluctuation-dissipation theorem.\cite{Brown1963} However, out of equilibrium
a proper microscopic derivation is required.
Microscopic derivations of the stochastic LLG equation
out of equilibrium, available in the literature, use the model of a
localized spin coupled to itinerant
electrons,\cite{Rebei2005,Katsura2006,Nunez2008,Kamenev2008}
or deal with non-interacting electrons.\cite{Foros2005}
In contrast to this approach, we start from a purely electronic system
where the magnetization arises as a consequence of the Stoner
instability. Our derivation has certain similarity with that of
Ref.~\onlinecite{Duine2007} for a bulk ferromagnet, where the direction
of magnetization is fixed and cannot be changed globally, so its local
fluctuations are small and their description by a gaussian action is
sufficient. This situation should be contrasted  to the case of a
nanoparticle
where the direction of the magnetization can be completely randomized by
the fluctuations, so that the effective action for the direction of the
classical magnetization has a non-gaussian part. The bias-driven
Brownian motion of the magnetization with a fixed direction (due to
and easy-axis anisotropy and ferromagnetic electrodes) has been also
studied in Ref.~\onlinecite{Waintal2003} using rate equations.

\begin{figure}
\epsfxsize=0.5\textwidth
\centerline{\epsfbox{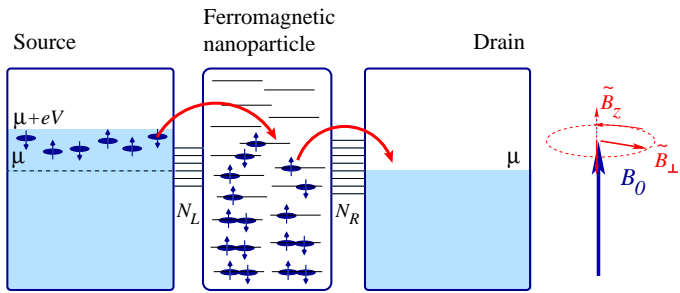}}
\caption{\label{fig:setup}(Color online).
Device setup considered in this work:
a small ferromagnetic particle (quantum dot) coupled to two
non-magnetic leads (see text for details).}
\label{fig:1}
\end{figure}

We assume that the single-electron spectrum of the particle, which
is also called a quantum dot in the literature, to be chaotic and
described by the random-matrix theory~\cite{BeenakkerRMP,ABG}. To
take into account the electron-electron interactions in the dot we
use the universal Hamiltonian,\cite{Kurland2000} with a generalized
spin part, corresponding to a ferromagnetic particle. Electrons
occupy the quantum states of the full Hamiltonian and form a net
spin of the particle of order of $S_0\gg 1$; throughout the paper we
use $\hbar =1$. The dot is coupled to two leads, see
Fig.~\ref{fig:setup}, which we assumed to be non-magnetic. The
approach can be easily extended to the case of magnetic leads. The
number~$N_{ch}$ of the transverse channels in the leads, which are
well coupled to the dot, is assumed to be large, $N_{ch}\gg{1}$.
Equivalently, the escape rate $1/\tau$ of electrons from the dot into the
leads is large compared to the single-electron mean level
spacing $\mls$ in the dot. This coupling to the leads is responsible
for tunnelling processes of electrons between states in the leads
and in the dot with random spin orientation. As a result of such
tunnelling events, the net spin of the particle changes.  We show
that this exchange of electrons gives rise both to the Gilbert
damping and the magnetization noise in the presented model, and
under conditions specified below, the time evolution of the particle
spin is described by the stochastic LLG equation.

We study in detail the conditions for applicability of the stochastic
approach. We find that these limits are set by three independent criteria.
First, the contact resistance should be low compared to the resistance
quantum, which is equivalent to $N_{ch}\gg{1}$. If this condition is broken,
the statistics of the noise cannot be considered gaussian. Physically,
this condition
means that each channel can be viewed as an independent source of noise,
so the contribution of many channels results in the gaussian noise by virtue
of the central limit theorem if $N_{ch}\gg{1}$. Second,
the system should not be too close to the Stoner instability: the mean-field
value of the total spin $S_0\gg\sqrt{N_{ch}}$. If this condition is violated,
the fluctuations of the absolute value of the magnetization become of
the order of the magnetization itself. Third, $S_0^2\gg{T}_\mathrm{eff}/\mls$,
where $T_\mathrm{eff}$ is the effective temperature of the system, which is
the energy scale of the electronic distribution function determined by a combination of
temperature and bias voltage (the Boltzmann constant $k_{\rm B}=1$
throughout the paper). Otherwise, the separation of the degrees of
freedom into slow (the direction of the magnetization) and fast (the electron
dynamics and the fluctuations of the absolute value of the magnetization)
is not possible.

In the present model we completely neglect the spin-orbit
interaction inside the particle, whose effect is assumed to be weak
as compared to the effect of the leads.\cite{Tserkovnyak2002} The
effects of the electron-electron interaction in the charge channel
(weak Coulomb blockade) are suppressed for $N_{ch}\gg{1}$,\cite{ABG}
so we do not consider it.

As an application of the formalism, we consider the magnetic
susceptibility in the ferromagnetic resonance measurements, which is
a standard characteristic of magnetic samples. Recently, a progress
was reported in measurements of the magnetic susceptibility on small
spatial scales in response to high-frequency magnetic
fields.\cite{Tamaru2002} Measurements of the ferromagnetic resonance
were also reported for nanoparticles, connected to leads for a
somewhat different setup in Ref.~\onlinecite{Sankey2006}.

The paper is organized as follows. In Sec.~\ref{sec:model} we introduce
the model for electrons in a small metallic particle subject to Stoner
instability. In Sec.~\ref{sec:action} we analyze the effective bosonic
action for the magnetization of the particle. In Sec.~\ref{sec:langevin}
we obtain the equation of motion for the magnetization with the stochastic
Langevin term, which has the form of the stochastic Landau-Lifshitz-Gilbert
equation, and derive the associated Fokker-Planck equation. In
Sec.~\ref{sec:applicability} we discuss the conditions for the applicability
of the approach. In Sec.~\ref{sec:susceptibility} we calculate the
magnetic susceptibility from the stochastic LLG equation.

\section{Model and basic formalism}\label{sec:model}

Within the random matrix theory framework, electrons in a closed
chaotic quantum dot are described by the following fermionic action:
\begin{widetext}
\begin{equation}
\mathcal{S}[\psi,\psi^*]=\oint{dt}
\left[\sum_{n,n'=1}^N
\psi^\dagger_{n}(t)\left(\delta_{nn'}i\partial_t-H_{nn'}\right)\psi_{n'}(t)
-E(\vec{S}(t))\right],\quad
S_i\equiv\sum_{n=1}^N\psi^\dagger_n\,\frac{\spinmat{\sigma}^i}2\,\psi_n\,.
\end{equation}
Here $\psi_n$ is a two-component Grassmann spinor, $t$~runs
along the Keldysh contour, as marked by~$\oint$;
$\spinmat{\sigma}^{x,y,z}$ are the Pauli matrices (we use the
hat to indicate matrices in the spin space and use the
notation~$\hat\sigma_0$ for the $2\times{2}$ unit matrix).
$H_{nn'}$ is an $N\times{N}$ random matrix from a gaussian
orthogonal ensemble, described by the pair correlators:
\begin{eqnarray}
&&\overline{\displaystyle H_{mn}H_{m'n'}}=\frac{N\mls^2}{\pi^2}
\left[\delta_{mn'}\delta_{nm'}+\delta_{mm'}\delta_{nn'}\right].
\end{eqnarray}
Here $\mls$~is the mean single-particle level spacing in the dot.

The magnetization energy $E(\vec{S})$ is the generalization of
the $J_sS^2$ term in the universal Hamiltonian for the electron-electron
interaction in a chaotic quantum dot.\cite{Kurland2000}
Since we are going
to describe a ferromagnetic state with a large value of the total
spin on the dot, we must go beyond the quadratic term; in fact,
all terms should be included. $E(\vec{S})$~can be viewed as
the sum of all irreducible many-particle vertices in the spin channel,
obtained after integrating out degrees of freedom with high energies
(above Thouless energy); the corresponding term in the action is
thus local in time, and can be written as the time integral of an
instantaneous function $E(\vec{S}(t))$. This functional can be
decoupled using the Hubbard-Stratonovich transformation with
a real vector field $\vec\magn(t)$, which we call below
the internal magnetic field:
\begin{equation}
\exp\left(-i \oint{dt} E(\vec{S})\right)=\int{\cal{D}}\vec\magn(t)\,
\exp\left(i\oint{dt}(2\vec\magn\cdot\vec{S}-
\tilde{E}(\vec\magn))\right).
\label{HubbardStr=}
\end{equation}
We rewrite the action $\mathcal{S}[\psi,\psi^*]$ in the form
\begin{equation}
\mathcal{S}[\psi,\psi^*,\vec\magn]=\oint{dt}
\left[\sum_{n,n'=1}^N\psi^\dagger_{n}(t)
\left(\spinmat{G}^{-1}\right)_{nn'}
\psi_{n'}(t)
-\tilde{E}(\vec\magn(t))\right],
\label{RMTAction2=}
\end{equation}
\end{widetext}
where the inverse Green's function
\begin{equation}\label{Gf=}
\left(\spinmat{G}^{-1}\right)_{nn'}=\left(i\spinmat{\sigma}_0\partial_t+\vec\magn\cdot\spinmat{\vec\sigma}\right)
\delta_{nn'}-{H}_{nn'}\spinmat{\sigma}_0
\end{equation}
is a matrix in time variables $t,t'$, in orbital indices $n$ and
$n'$ with $1\leq n,n' \leq N$, in spin indices, and
in forward~($+$) and backward~($-$) directions on the Keldysh contour.
Integration over fermionic fields $\psi_n,\psi_n^\dagger$ yields the purely bosonic action:
\begin{equation}
\mathcal{S}[\vec\magn]=
-i\tr\left\{\ln(-i\spinmat{G}^{-1})\right\}
-\oint{d}t\,\tilde{E}(\vec\magn(t)),
\label{bosonicaction=}
\end{equation}
where the trace is taken over \emph{all} indices of the Green's
function, listed above.

In the space of forward and backward directions on the Keldysh contour,
we perform the standard Keldysh rotation, introducing
the retarded~($G^R$), advanced~($G^A$), Keldysh~($G^K$), and zero~($G^Z$)
components of the Green's function:
\begin{equation}
\left(\begin{array}{cc} \spinmat G^R & \spinmat G^K \\ \spinmat G^Z & \spinmat G^A \end{array}\right)
=\frac{1}{2}
\left(\begin{array}{cc} 1 & 1 \\ 1 & -1 \end{array}\right)\!\!
\left(\begin{array}{cc} \spinmat G^{++} & \spinmat G^{+-} \\ \spinmat G^{-+} & \spinmat G^{--}
\end{array}\right)\!\!
\left(\begin{array}{cc} 1 & 1 \\ -1 & 1 \end{array}\right)\! ,
\end{equation}
as well as the classical~($\vec\magn^{cl}$) and quantum~($\vec\magn^q$)
components of the field:
\begin{equation}
\left(\begin{array}{cc} \vec\magn^{cl} & \vec\magn^q \\
\vec\magn^q & \vec\magn^{cl} \end{array}\right)
=\frac{1}{2}
\left(\begin{array}{cc} 1 & -1 \\ 1 & 1 \end{array}\right)
\left(\begin{array}{cc} \vec\magn^+ & 0 \\ 0 & -\vec\magn^- \end{array}\right)
\left(\begin{array}{cc} 1 & 1 \\ 1 & -1 \end{array}\right).
\end{equation}
We will also write this matrix as
$\vec\magn=\vec\magn^{cl}\tau^{cl}+\vec\magn^q\tau^q$,
where $\tau^{cl}$~and~$\tau^q$ are  $2\times{2}$ matrices in
the Keldysh space coinciding with the unit $2\times{2}$ matrix and the first Pauli matrix,
$\hat \sigma_x$, respectively.

The saddle point of the bosonic action \req{bosonicaction=} is found
by the first order variation with respect to~$\vec\magn^{cl,q}(t)$, which gives the
self-consistency equation:
\begin{equation}
\vec\magn^q(t)=0\,,\quad
-\frac{\partial\tilde{E}(\vec\magn^{cl}(t))}{\partial\magn_j^{cl}(t)}
=\frac{i}2\tr_{n,\sigma}
\left\{\spinmat{\sigma}^j\spinmat{G}_{nn}^K(t,t)\right\}.
\label{selfconsG=}
\end{equation}
We also
note that the right-hand side of this equation is proportional
to the total spin of electrons of the particle
for a given trajectory of~$\vec\magn^{cl}(t)$:
\begin{equation}
\vec{S}(t)=\frac{i}4\tr_{n,\sigma}
\left\{\spinmat{\sigma}^j\spinmat{G}^K(t,t)\right\}.
\label{S_t=}
\end{equation}
In \reqs{selfconsG=} and \rref{S_t=}, the trace is taken over orbital and spin indices only.

In the limit $N\to\infty$, one can obtain a closed equation for the
Green's function traced over the orbital indices:\cite{Ahmadian2005}
\begin{equation}
\hat{g}(t,t')=\frac{i\mls}\pi\sum_{n=1}^N\hat{G}_{nn}(t,t').
\label{smallg=}
\end{equation}
The matrix $\hat{g}(t,t')$ satisfies the following constraint:
\begin{equation}
\int\hat{g}(t,t'')\,\hat{g}(t'',t')\,dt''=\tau^{cl}\hat\sigma_0\delta(t-t'),
\end{equation}
where the right-hand side is just the direct product of unit matrices
in the spin, Keldysh, and time indices. The Wigner transform of
$\hat{g}^K(t,t')$ is related to the spin-dependent distribution
function $\hat{f}(\ep,t)$ of electrons in the dot:
\begin{equation}
\int\limits_{-\infty}^\infty\hat{g}^K(t+\tilde{t}/2,t-\tilde{t}/2)\,
e^{i\ep\tilde{t}}\,d\tilde{t}=2\hat{f}(\ep,t).
\end{equation}
In equilibrium, $\hat{f}(\ep)=\hat\sigma_0\tanh(\ep/2T)$.

The self-consistency condition~(\ref{selfconsG=}) takes the form
\begin{equation}
-\frac{\partial\tilde{E}(\vec\magn^{cl}(t))}{\partial\magn_i^{cl}(t)}
=\frac\pi{2\mls}\lim_{t'\to{t}}\tr_{\sigma}
\left\{\spinmat{\sigma}^i\spinmat{g}^K(t,t')\right\}
-\frac{2\magn_i^{cl}(t)}{\mls}. \label{selfconsanom=}
\end{equation}
The last term takes care of the anomaly arising from
non-commutativity of the limits $N\to\infty$ and $t'\to{t}$.

In this paper we consider the dot coupled to two leads, identified as left~($L$)
and right~($R$). The leads  have $N_L$~and~$N_R$
transverse channels, respectively, see Fig~\ref{fig:1}.  For non-magnetic
leads and spin-independent  coupling between the leads and the particle,
we can characterize each channel by its transmission~$T_n$ with
$0<{T}_n\leq{1}$ and by the distribution function of electrons
in the channel~$\df_n(t-t')$, assumed to be stationary.
We consider the limit of strong coupling between the leads and the
particle, $\sum_{n=1}^{N_{ch}} T_n\gg 1$.

The coupling to the leads gives rise to a self-energy term, which
should be included in the definition of the Green's function,
Eq.~(\ref{Gf=}). Without going into details of the derivation,
presented in Ref.~\onlinecite{Ahmadian2005}, we give the final form
of the equation for the Green's function traced over the orbital
states, \req{smallg=}:
\begin{equation}\begin{split}\label{openUsadel=}
&\left[\partial_t-i\vec\magn\cdot\spinmat{\vec{\sigma}},\spinmat{g}\right]
\\
& = \sum_{n=1}^{N_{ch}}\frac{T_n\mls}{2\pi}
\left(\begin{array}{cc}-\mathcal{F}_n\spinmat{g}^Z &
\spinmat{g}^R\mathcal{F}_n-\mathcal{F}_n\spinmat{g}^A-\spinmat{g}^K\\
\spinmat{g}^Z & -\spinmat{g}^Z\mathcal{F}_n
\end{array}\right)
\\
&\times
\left[\unitmatrix+\frac{T_n}2\left(\begin{array}{cc}
\spinmat{g}^R-\unitmatrix+\mathcal{F}_n\spinmat{g}^Z &
\spinmat{g}^R\mathcal{F}_n+\mathcal{F}_n\spinmat{g}^A \\
0 & -\spinmat{g}^A-\unitmatrix+\spinmat{g}^Z\mathcal{F}_n
\end{array}\right)\right]^{-1}.
\end{split}
\end{equation}
Here the products of functions include convolution in time variables.
This equation is analogous to the Usadel equation used
in the theory of dirty superconductors.\cite{Usadel1970}

To conclude this section, we discuss the dependence
$\tilde{E}(\vec\magn)$. Deep in the ferromagnetic state, \emph{i.e.}
far from the Stoner critical point, we expect the mean-field
approach to give a good approximation for the total spin
of the dot. Namely, the mean field acting on
the electron spins, is given by $2\magn_0=dE(S)/dS\equiv{E}'(S)$.
We then require that the response of the system to this field gives
the same average value for the spin:
\begin{equation}
S_0=\frac{2\magn_0}{2\mls}=\frac{E'(S_0)}{2\mls}.
\end{equation}
Here we evaluated $S_0$ from \req{S_t=} and applied the self-consistency
equation \rref{selfconsanom=} to equilibrium state with
$\spinmat{g}^K\propto\spinmat{\sigma}_0$,
when the contribution of the first term in the right hand side of
\req{selfconsanom=} vanishes.

Not expecting strong deviations of the magnitude of the spin from
the mean-field value, we focus on the form of $\tilde{E}(\vec\magn)$
when $|\vec\magn|\approx\magn_0$. The inverse Fourier transform of
Eq.~(\ref{HubbardStr=}) and angular integration for the isotropic~$E(S)$
gives
\begin{equation}
e^{-i\tilde{E}(\magn)\Delta t}=\mathrm{const}\int\limits_0^\infty
\frac{\sin{2S\magn\Delta t}}{2S\magn\Delta t}\,e^{-iE(S)\Delta t}
S^2\,dS,
\end{equation}
where $\Delta t$ is the infinitesimal time increment used in the
construction of the functional integral in \req{HubbardStr=}.

Expanding $E(S)$ near the mean-field value~$S_0$,
\begin{equation}
E(S)\approx{E}(S_0)+{E}'(S-S_0)+\frac{E''}2\,(S-S_0)^2,
\end{equation}
performing the integration in the stationary phase approximation and
using $S_0=\magn_0/\mls=-E'/(2\mls)$,
we obtain
\begin{equation}
\tilde{E}(\magn)=-2 \frac{(\magn-\magn_0-E''S_0/2)^2}{E''}+\tilde E_0,
\label{tildeE=}
\end{equation}
where $\tilde E_0$ is $\magn-$independent term. This expression for
$\tilde E(\vec{\magn})$ defines the action $\mathcal{S}[\vec\magn]$,
\req{bosonicaction=}.

The energy $E(S)$ does not contain the energy $E_{\rm orb}(S)$,
associated with the orbital motion of electrons in the particle.
Namely, to form a total spin $S$ of the particle, we have to
redistribute $S$ electrons over orbital states, which changes the
orbital energy of electrons by $E_{\rm orb}(S)\simeq \mls S^2$. The total
energy $E_{\rm tot}(S)$ of the particle is the sum of two terms:
$E_{\rm tot}(S) = E(S)+E_{\rm orb}(S)$.  Similarly, we obtain the total
energy of the system in terms of internal magnetic field
\begin{equation}
\begin{split}
\tilde{E}_{\mathrm{tot}}(\magn) & = \tilde{E}(\magn)-\frac{\magn^2}\mls
\\
&=
-2\left(\frac{1}{E''}+\frac{1}{2\mls}\right)(\magn-\magn_0)^2
+\tilde E_1,
\label{tildeEtot=}
\end{split}
\end{equation}
where $\tilde E_1$ does not depend on $\magn$. We notice that the
extremum of $\tilde{E}_{\mathrm{tot}}(\magn)$ corresponds to
$\magn=\magn_0$ and describes the expectation value of the internal
magnetic field in an isolated particle. The energy cost of
fluctuations of the magnitude of the internal magnetic field is
characterized by the coefficient $1/E''+1/2\mls$.

\section{Keldysh action}\label{sec:action}

In this Section we analyze the action \req{bosonicaction=} for the internal magnetic field
$\vec{\magn}^\alpha$.
We expect that the classical
component $\vec{\magn}^{cl}(t)$ of this field contains fast and
small oscillations of its magnitude around the mean-field value
$h_0$. We further expect that the orientation of $\vec{\magn}^{cl}(t)$
changes slowly in time, but is not restricted to small deviations
from some specific direction. Based on this picture,
we introduce a unit vector~$\vec{n}(t)$,
assumed to depend slowly on time, and write
\begin{equation}
\vec{\magn}^{cl}(t)=(\magn_0+ \magn_\|^{cl}(t))\vec{n}(t)\,,
\label{hcl=}
\end{equation}
where $ \magn_\|^{cl}(t)$ is assumed to be fast  and small.
We expand the action~(\ref{bosonicaction=}) to the second order in
small fluctuations of the quantum component $\vec{\magn}^{q}(t)$ and
the radial classical component $\magn_\|^{cl}(t)$:
\begin{eqnarray}
\mathcal{S}[\vec\magn]&\approx&-\frac{2\pi}{\mls}\int{d}t
 \vec{g}^K(t,t)\vec{\magn}^q(t)
+\nonumber\\
&&{}+\frac{8}{E''}\int{d}t \magn_\|^{cl}(t)\vec{n}(t)\vec{\magn}^q(t)
\nonumber\\
&&{}-\int{d}t\,{d}t'\,\Pi^R_{ij}(t,t')\,
\magn_i^q(t)\,\magn_\|^{cl}(t')\,n_j(t') \nonumber\\
&&{}-\int{d}t\,{d}t'\,\Pi^A_{ij}(t,t')\,
\magn_\|^{cl}(t)\,n_i(t)\,
\magn_j^q(t') \nonumber\\
&&{}-\int{d}t\,{d}t'\,\Pi^K_{ij}(t,t')\,
\magn_i^q(t)\,\magn_j^q(t')\,.\label{claction=}
\end{eqnarray}
The applicability of this quadratic expansion is discussed in
Sec.~\ref{sec:gaussian}.

In \req{claction=} we introduced the polarization operator,
defined as the kernel of
the quadratic part of the action of the fluctuating bosonic fields:
\begin{subequations}\begin{eqnarray}
&&\left(\begin{array}{cc} \Pi^Z & \Pi^A \\ \Pi^R & \Pi^K
\end{array}\right)
\equiv
\left(\begin{array}{cc} \Pi^{cl,cl} & \Pi^{cl,q} \\
\Pi^{q,cl} & \Pi^{q,q} \end{array}\right)
,\\
&&\Pi_{ij}^{\alpha\beta}(t,t')=\frac{i}2\,
\frac{\delta^2\Tr{\ln{G}^{-1}}}
{\delta\magn^\beta_j(t')\,\delta\magn^\alpha_i(t)}\,,
\label{definePi=}
\end{eqnarray}\end{subequations}
where $\alpha,\beta=cl,q$ and $i,j=x,y,z$.
The short time anomaly is explicitly taken into account in
the definition of the polarization operators, see \req{Pi=} below.

The first term of \req{claction=} contains the vector
$\vec{g}^K$ of the Keldysh component of the Green function
$\spinmat{g}^K=\spinmat{\sigma}_0g^K_0
+\spinmat{\vec\sigma}\cdot\vec{g}^K$.
We emphasize that the Green's function and the polarization
operator in \req{claction=} are calculated at $\magn_\|^{cl}(t)=0$
and $\vec{\magn}^q(t)=0$
for a given trajectory of the classical field
$h_0\vec{n}(t)$.

\subsection{Keldysh component of the Green function}

For the Green's function in the classical field we have
\begin{equation}
\spinmat{g}^R(t,t')=-\spinmat{g}^A(t,t')=\spinmat{\sigma}^0\delta(t-t'),
\label{gRA0=}
\end{equation}
while the Keldysh component satisfies the equation
\begin{equation}
\left[\partial_t-i\magn_0\vec{n}\cdot\spinmat{\vec{\sigma}},\spinmat{g}^K\right]=
\sum_{n=1}^{N_{ch}}\frac{T_n\mls}{2\pi}
\left(2\df_n-\spinmat{g}^K\right).
\end{equation}
We introduce the notation
\begin{equation}
\frac{1}\tau=\sum_{n=1}^{N_{ch}}\frac{T_n\mls}{2\pi}
=\frac{1}{\tau_L}+\frac{1}{\tau_R},
\end{equation}
Then the scalar $g^K_0$ and vector $\vec{g}^K$
components of $\spinmat{g}^K=\spinmat{\sigma}_0g^K_0
+\spinmat{\vec\sigma}\cdot\vec{g}^K$ satisfy two coupled equations:
\begin{subequations}\begin{eqnarray}
&&\left[\partial_{t}+\partial_{t'}+\frac{1}\tau\right]g_0^K(t,t')=
\sum_{n=1}^{N_{ch}}\frac{T_n\mls}{2\pi}\,2\df_n(t-t')
\nonumber\\
&&+i\magn_0[\vec{n}(t)-\vec{n}(t')]\cdot\vec{g}^K(t,t'),
\label{scalgKtt=}\\
&&\left[\partial_{t}+\partial_{t'}+\frac{1}\tau\right]\vec{g}^K(t,t')+
\magn_0[\vec{n}(t)+\vec{n}(t')]\times\vec{g}^K(t,t')
\nonumber\\&&=
i\magn_0[\vec{n}(t)-\vec{n}(t')]\,{g}^K_0(t,t').
\label{vecgKtt=}
\end{eqnarray}\end{subequations}

As a zero approximation, we can consider the stationary situation:
$g_0^K(t,t')=g_0^K(t-t')$ and $\vec{n}(t)=\mathrm{const}$. In this case,
we have
\begin{equation}
g_0^K(t,t')=\frac{\tau}{\tau_L}\,2\df_L(t-t')+\frac{\tau}{\tau_R}\,2\df_R(t-t')
\label{gK0st=}
\end{equation}
and $\vec{g}^K=0$.

For an arbitrary time dependence $\vec{n}(t)$, Eq.~(\ref{vecgKtt=}) cannot
be solved analytically.
However, if the variation of $\vec{n}(t)$ is slow enough,
we can make a gradient expansion:
\begin{equation}
\left(\partial_{\overline{t}}+\frac{1}\tau\right)\vec{g}^K
+2\magn_0\vec{n}\times\vec{g}^K
=i\tilde{t}\magn_0\dot{\vec{n}}g_0^K.
\label{eqforgk=}
\end{equation}
Here we introduced $\overline{t}=(t+t')/2$, $\tilde{t}=t-t'$,
$\partial_{t}+\partial_{t'}=\partial_{\overline{t}}$. The
dependence on~$\tilde{t}$ is split off and remains unchanged, while for the
dependence on~$\overline{t}$ the solution is determined by a linear
operator~$\mathcal{L}^+_{\vec{n}}$:
\begin{subequations}\begin{align}\label{lpm=}
&\mathcal{L}^\pm_{\vec{n}}=
\left(
\frac{1}\tau\pm\partial_{\overline{t}}\pm 2\magn_0\vec{n}\times{}\right)^{-1},\\
&\mathcal{L}^+_{\vec{n}}(\omega)\,\vec{X}(\omega)=
\frac{\vec{n}(\vec{n}\cdot\vec{X}(\omega))}{-i\omega+1/\tau}
+\nonumber\\ &\qquad{}
+\frac{1}2\sum_\pm\frac{-\vec{n}\times[\vec{n}\times\vec{X}(\omega)]
\pm{i}[\vec{n}\times\vec{X}(\omega)]}
{-i(\omega\pm{2}\magn_0)+1/\tau}\,.
\end{align}\end{subequations}
Here we assume that the direction of the internal magnetic field
$\vec{n}$ changes slowly in time, and $|\dot{\vec{n}}|\tau\ll{1}$.

Thus, all perturbations of $\vec{g}^K$ decay with the characteristic time~$\tau$.
In particular, the solution of \req{eqforgk=} has the form
\begin{equation}
\vec{g}^K=i\tilde{t}\magn_0\mathcal{L}^+_{\vec{n}}\dot{\vec{n}}g_0^K
\approx{i}\tilde{t}\magn_0\tau\,g_0^K\,
\frac{\dot{\vec{n}}+2\magn_0\tau\,\vec{n}\times\dot{\vec{n}}}
{(2\magn_0\tau)^2+1}. \label{gkt0=}
\end{equation}

Expression for the first term in \req{claction=} can be easily
obtained from \req{gkt0=} by taking the limit $\tilde t\to 0$ and
taking into account that any fermionic distribution function in the
time representation has the following equal-time asymptote:
\begin{equation}
g_0^K(t,t')\approx\frac{2}{i\pi}\frac{1}{t-t'},\quad(t\to{t}').
\label{gk0tt=}
\end{equation}
We have
\begin{equation}
\vec{g}^K(t,t)=\frac{2 \magn_0\tau}{\pi}\,
\frac{\dot{\vec{n}}+2\magn_0\tau\,\vec{n}\times\dot{\vec{n}}}
{(2\magn_0\tau)^2+1}.
\label{vecgKtt1=}
\end{equation}
We notice that $\vec{n}\cdot \vec{g}^K(t,t)=0$, and therefore the
first term in the action \req{claction=} is coupled only to the
tangential fluctuations of $\vec{\magn}^q(t)\perp \vec{n}(t)$.

\subsection{Polarization operator}
We express the polarization operator in terms of the unit
vector $\vec{n}(t)$. The polarization operator can be represented
as the response of the Green's functions to a change in the field, as follows
directly from the definition~(\ref{definePi=}) and
the expression~(\ref{bosonicaction=}) for the action:
\begin{equation}
\Pi_{ij}^{\alpha\beta}(t,t')=
\frac{\pi}{2\mls}\,\frac{\tr\nolimits_{4\times{4}}
\left\{\tau^\alpha
\spinmat{\sigma}^i\delta\spinmat{g}(t,t)\right\}}{\delta\magn^\beta_j(t')}
-\frac{2}\mls\,\tau^q_{\alpha\beta}\delta_{ij}\delta(t-t').
\label{Pi=}
\end{equation}
Here the Green function $\delta \spinmat{g}(t,t)$ can be calculated as the
first-order response of the solution of \req{openUsadel=} to small arbitrary
(in all three directions) increments of $\delta\vec\magn^{cl}(t)$
and $\delta\vec\magn^q(t)$.
The zero-order solution of \req{openUsadel=} in the
field $\vec\magn^{cl}=\magn_0\vec{n}$ and $\vec\magn^q=0$ is
\begin{equation}
\hat{g}(t,t)=\hat\sigma_0\left(\begin{array}{cc}
\delta(t-t') & g_0^K(t-t') \\ 0 & -\delta(t-t') \end{array}\right).
\end{equation}

First, we calculate $\delta\spinmat{g}^Z$, which responds only
to~$\delta\vec\magn^q$:
\begin{equation}
\bes
\left[\partial_{t}+\partial_{t'}-\frac{1}\tau\right]\delta\spinmat{g}^Z(t,t')
-i\magn_0 & \spinmat{\vec\sigma}\cdot\vec{n}(t)\,\delta\spinmat{g}^Z(t,t')
\\ +
\delta\spinmat{g}^Z(t,t')\,i\magn_0\spinmat{\vec\sigma}\cdot\vec{n}(t')&
=2i\spinmat{\vec\sigma}\cdot\delta\vec\magn^q(t)\,\delta(t-t').
\end{split}
\end{equation}
Since $\partial_t+\partial_{t'}=\partial_{\overline{t}}$, the solution always remains
proportional to~$\delta(t-t')$:
\begin{equation}
\delta\spinmat{g}^Z(t,t')=
-2i\spinmat{\vec\sigma}(\mathcal{L}_{\vec{n}}^-\delta\vec\magn^q)(t)\,\delta(t-t').
\label{deltagZ=}
\end{equation}
Given $\delta\spinmat{g}^Z$, components
$\delta\spinmat{g}^{R,A}$ can be found either from Eq.~(\ref{openUsadel=}),
or, equivalently, using the constraint $\spinmat{g}^2=\unitmatrix$:
\begin{equation}
\spinmat{g}\,\delta\spinmat{g}+\delta\spinmat{g}\,\spinmat{g}=0
\;\;\;\Rightarrow\;\;\;
\delta\spinmat{g}^R=-\frac{\spinmat{g}^K\delta\spinmat{g}^Z}2,
\;\;\;
\delta\spinmat{g}^A=\frac{\delta\spinmat{g}^Z\,\spinmat{g}^K}2.
\end{equation}
We notice that both $\delta\spinmat{g}^{R,A}$ respond only to
$\vec{\magn}^q(t)$ and, therefore,
\begin{equation}
\Pi^Z_{ij}(t,t')\propto
\frac{\tr\{\spinmat{\sigma}^i(\delta \spinmat{g}^R(t,t)+\delta \spinmat{g}^A(t,t))\}}
{\delta\magn^{cl}_j(t')}\equiv 0.
\end{equation}
This equation ensures that the action along the Keldysh contour
vanishes for $\vec{h}^q\equiv 0$.

To evaluate the remaining three components of the polarization
operator, we can apply the variational derivatives to the sum of
$\delta \spinmat{g}^K(t,t)+\delta \spinmat{g}^Z(t,t)$ with respect
to either classical $\delta\vec{\magn}^{cl}(t')$ or quantum
$\delta\vec{\magn}^{q}(t')$ field, which give $\Pi^R_{ij}(t,t')$ and
$\Pi^K_{ij}(t,t')$, respectively. Then, the advanced component
$\Pi^A_{ij}(t,t')=[\Pi^R_{ji}(t',t)]^*$.

The equation for~$\delta\spinmat{g}^K=\spinmat{\vec{\sigma}}\cdot\delta\vec{g}^K$ reads as
\begin{subequations}
\begin{equation}
\begin{split}
&\left[\partial_{t}+\partial_{t'}+\frac{1}\tau\right]\delta\vec{g}^K(t,t')\\
& \quad\quad +h_0\left[
\vec{n}(t)\times\delta \vec{g}^K(t,t')-\delta \vec{g}^K(t,t')\times\vec{n}(t')
\right]\\
 & =i\left[\delta\vec{\magn}^{cl}(t)-\delta\vec{\magn}^{cl}(t')\right] g_0^K(t-t')
\\
&\quad\quad
-2i\delta\vec{\magn}^q(t)\delta(t-t')-\vec{Q}(t,t')
,\label{deltagK=}
\end{split}
\end{equation}
where
\begin{equation}
\begin{split}
\vec{Q}(t,t')&=\sum_{n=1}^{N_{ch}}
\frac{T_n\mls}{2\pi}\left(\frac{g^K_0}{2}\,\delta\vec{g}^Z\,\frac{g^K_0}2
+\mathcal{F}_n\,\delta\vec{g}^Z\mathcal{F}_n\right)
\\
&
-\sum_{n=1}^{N_{ch}}\frac{T_n(1-T_n)\mls}{2\pi}\,\left(\frac{g^K_0}2-\mathcal{F}_n\right)\delta\vec{g}^Z
\left(\frac{g^K_0}2-\mathcal{F}_n\right)
\end{split}
\end{equation}
\end{subequations}
and $\delta\vec{g}^Z=\tr\{\spinmat{\vec{\sigma}}\delta \spinmat{g}^Z\}/2 $
with $\delta \spinmat{g}^Z$ given by \req{deltagZ=}.

To calculate the retarded component $\Pi_{ij}^R$ of the polarization
operator, we calculate the response of $\delta\vec{g}^K(t,t')$ to
$\delta\vec{\magn}^q$ in the limit $t'\to t$. Using the asymptotic
behavior of the Fermi function, \req{gk0tt=},
we obtain:
\begin{equation}
\delta\vec{g}^K(t,t)=\int\frac{d\omega}{2\pi}\,
\frac{-2i\omega}{\pi}\,\mathcal{L}^+_{\vec{n}}(\omega)\,
\delta\vec\magn^{cl}(\omega)\,e^{-i\omega{t}}.
\end{equation}
Substituting this expression for $\delta\vec{g}^K(t,t)$ to
\req{Pi=}, we obtain
\begin{subequations}
\begin{eqnarray}
\Pi_{ij}^R(\omega) & = &
\Pi_{\|,ij}^R(\omega)+\Pi_{\perp,ij}^R(\omega)\,,
\end{eqnarray}
with
\begin{eqnarray}
\Pi_{\|,ij}^R(\omega) & = & -\frac{2}\mls\,
\frac{n_in_j}{1-i\omega\tau}\,,\label{PRrad=}
\\
\Pi_{\perp,ij}^R(\omega) & = & -\frac{2}{\mls}
\sum_\pm
\frac{\delta_{ij}-n_in_j\pm{i}e_{ijk}n_k}{2}
\\
&&
\times\frac{(1\pm{2}i\magn_0\tau)}{1-i(\omega\mp{2}\magn_0)\tau}\,.\nonumber
\end{eqnarray}
\end{subequations}
Here we represented the polarization operator $\Pi_{ij}^R(\omega)$
as a sum of the radial, $\Pi_{\|,ij}^R(\omega)$, and tangential, $\Pi_{\perp,ij}^R(\omega)$,
terms. We note that the action
\req{claction=} contains only the radial component of the retarded
and advanced polarization operators because we do not perform
expansion in terms of the tangential fluctuations of the classical
component of the field $\vec{h}^{cl}(t)$.

In response to $\delta\vec\magn^q$, both  corrections
$\delta\vec{g}^K(t,t')$ and $\delta\vec{g}^Z(t,t')$ contain terms
$\propto\delta(t-t')$,
However, their sum $\delta\vec{g}^K(t,t')+\delta\vec{g}^Z(t,t')$ remains finite
in the limit $t\to{t}'$:
\begin{widetext}\begin{subequations}\label{gK+gZ=}\begin{align}
\delta\vec{g}^K(t,t')+\delta\vec{g}^Z(t,t') & =
-2i\int{d}t''\int{d}t_1\,dt_2\,
\mathcal{L}_{\vec{n}}^+(\bar t-t_1)\,
\mathcal{Q}(t_1-t_2+\tilde{t}/2;t_2-t_1+\tilde{t}/2)\,
\mathcal{L}_{\vec{n}}^-(t_2-t'')\,\delta\vec\magn^q(t''),\\
\mathcal{Q}(\tau_1;\tau_2) & = \frac{2}{\tau}\delta(\tau_1)\delta(\tau_2)-\sum_{n=1}^{N_{ch}}\frac{T_n\mls}{2\pi}
\left[\frac{g^K_0(\tau_1)}2\,\frac{g^K_0(\tau_2)}2
+\mathcal{F}_n(\tau_1)\,\mathcal{F}_n(\tau_2)\right]\\
&
-\sum_{n=1}^{N_{ch}}
\frac{T_n(1-T_n)\mls}{2\pi}\left[\frac{g^K_0(\tau_1)}2-\mathcal{F}_n(\tau_1)\right]
\left[\frac{g^K_0(\tau_2)}2-\mathcal{F}_n(\tau_2)\right]\nonumber
\end{align}\end{subequations}\end{widetext}
with $\bar t=(t+t')/2$ and $\tilde t=t-t'$.

From \req{gK+gZ=} we obtain the following expression for the
Keldysh component of the polarization operator:
\begin{subequations}
\begin{align}
\Pi_{ij}^K(\omega) & = \Pi_{\|,ij}^K(\omega) +
\Pi_{\perp,ij}^K(\omega),
\\
\Pi_{\|,ij}^K(\omega)  & = -i\frac{n_in_j}{\omega^2+1/\tau^2}
\mathcal{R}(\omega),\\
\Pi_{\perp,ij}^K(\omega) & =
-\frac{i}2\sum_\pm
\frac{\delta_{ij}-n_in_j\pm{i}e_{ijk}n_k}
{(\omega\mp{2}\magn_0)^2+1/\tau^2}\mathcal{R}(\omega).
\label{PKperp=}
\end{align}\end{subequations}
Here function $\mathcal{R}(\omega)$ coincides with the noise power of
electric current through a metallic particle in the approximation of
non-interacting electrons
\begin{equation}
\begin{split}
&\mathcal{R}(\omega)=\sum_{n=1}^{N_{ch}}\int\frac{d\ep}{8\pi}T_n\\
&\times \Big{\{}
\left[8- g^K_0(\ep)\,g^K_0(\ep+\omega)
-4 \mathcal{F}_n(\ep)\,\mathcal{F}_n(\ep+\omega)\right]\\
&+(1-T_n)
\left[g^K_0(\ep)-2\mathcal{F}_n(\ep)\right]
\left[g^K_0(\ep+\omega)-2\mathcal{F}_n(\ep+\omega)\right]\Big{\}}.
\label{calR=}
\end{split}
\end{equation}
In principle, electron-electron
interaction in the charge channel can be taken into account. The
interaction modifies the expression \req{calR=} for
$\mathcal{R}(\omega)$ to the higher order\cite{CV07} in
$\tau\mls\ll 1$ and we neglect this correction here.

In this paper we consider a particle connected to electron leads at temperature $T$ with
the applied bias $V$. In this case, $\mathcal{F}_{L,R}(\ep)=\tanh(\ep-\mu_{L,R})/(2T)$
with $\mu_L-\mu_R=V$, and
the integration over~$\ep$ gives
\begin{equation}
2\pi\tau\mathcal{R}(\omega)=
4\omega\coth\frac\omega{2T}+\Xi \Upsilon_T(V,\omega),
\label{R=}
\end{equation}
where
\begin{equation}
\Upsilon_T(V,\omega)\equiv
\sum_\pm 2(\omega\pm{V})\coth\frac{\omega\pm{V}}{2T}
-4\omega\coth\frac\omega{2T}
\end{equation}
and $\Xi$ is
the "Fano factor" for a dot
\begin{equation}
\begin{split}
\Xi=\frac{\tau^2}{\tau_L\tau_R} & + \frac{\tau^3 \mls}{2\pi\tau_R^2}\sum_{n\in{L}}T_n(1-T_n)\\
&+
\frac{\tau^3 \mls}{2\pi\tau_L^2}\sum_{n\in{R}}T_n(1-T_n).
\end{split}
\end{equation}

At $|V|\gg{T}$ the function $\Upsilon_T(V,\omega)$ has two scales
of $\omega$: (i)~$T$~smears the non-analyticity at $\omega\to{0}$,
but the value of $\Upsilon_T(V,\omega)$ deviates from
$\Upsilon_T(V,0)$ at $|\omega|\sim|V|$. Thus, the typical time
scale above which one can approximate $\Pi^K(\omega)$ by a
constant is at least $\omega \ll \max\{T,|V|\}$.
In the limit $\omega\to{0}$ we have
\begin{equation}
\Pi^K_{ij}(\omega=0)  = -i\,\frac{8\tau{T}_{\mathrm{eff}}}{\mls}
\left(n_in_j+\frac{\delta_{ij}-n_in_j}{(2\magn_0\tau)^2+1}\right).
\label{Pklowfr=}
\end{equation}
The effective temperature ${T}_{\mathrm{eff}}$ is given by
\begin{equation}
{T}_{\mathrm{eff}}  \equiv  T+\Xi
\left(\frac{V}2\coth\frac{V}{2T}-T\right).
\label{Teff=}
\end{equation}

\subsection{Final form of the action}

We can rewrite the action for magnetization field
$\vec{h}=\{\vec{h}^{cl};\vec{h}^q\}$ with $\vec{h}^{cl}$ in the form of \req{hcl=}
as a sum of the radial and tangential terms:
\be
\mathcal{S}[\vec{h}]=\mathcal{S}_\|[\magn_\|^{cl},h^q_\|]+\mathcal{S}_\perp[\vec{n}(t),\vec{h}^q_\perp].
\label{actiontot=}
\ee
The radial term in the action has the form
\be
\begin{split}
& \mathcal{S}_\| [h_\|^{cl},h^q_\| ]=
(\mathcal{D}^{-1}_\|)^{\alpha\beta}(t,t')\, \magn_\|^{\alpha}(t)\,
\magn_\|^{\beta}(t'), \label{actionrad=}
\end{split}
\ee
where the inverse
function of the internal magnetic field propagator is given by
\be
\begin{split}(\mathcal{D}^{-1}_\|)^{\alpha\beta}(t,t')= &
\frac{4}{E''}\left(%
\begin{array}{cc}
  0  & 1  \\ 1  &  0 \\
\end{array}%
\right)\delta(t-t')
\\
&
-\left(
\begin{array}{cc}
  0                            & \Pi^R_\|(t,t')   \\ 
  {\Pi^A_\|(t,t')} & \Pi^K_\|(t,t')
\end{array}%
\right)\end{split}
\ee
and
$\Pi^{\alpha\beta}_\|(t,t')=n_in_j\Pi^{\alpha\beta}_{\|,ij}(t,t')$.
From this equation we find
\be
\mathcal{D}^R_{\|}(\omega)= \mathcal{D}^{q,cl}_{\|}(\omega)=
\frac{E''}{4}\,\frac{-i\omega+1/\tau}
{-i\omega+(\mls+E''/2)/(\tau\mls)},
\label{DparrR=}
\ee
and $\mathcal{D}^A_{\|}(\omega)=[\mathcal{D}^R_{\|}(\omega)]^*$. The
Keldysh component is
\be
\mathcal{D}^K_{\|}(\omega)=\mathcal{D}^{q,q}_{\|}(\omega)=
\mathcal{D}^R_{\|}(\omega) \Pi^K_\|(\omega)
\mathcal{D}^A_{\|}(\omega).
\label{DparrK=}
\ee

The tangential term in the action is
\be
\begin{split}
\mathcal{S}_\perp[\vec{n}(t),\vec{h}^q_\perp]&=
-\frac{4\magn_0\tau}{\mls}\int dt
\frac{(\dot{\vec{n}}+2\magn_0\tau\,\vec{n}\times\dot{\vec{n}})\vec{h}^q_\perp}
{(2\magn_0\tau)^2+1}\\
&-\frac{4}{\mls}\int dt \left[\vec{n}(t)\times[\vec{n}(t)\times\vec{B}]\right]\cdot \vec{h}^q_\perp(t)\\
&-\int{d}t\,{d}t'\,\magn_{\perp,i}^q(t) \Pi^K_{\perp,ij}(t-t')\,
\,\magn_{\perp,j}^q(t')\,.\label{actionperp=}
\end{split}
\ee
Here we recovered  the external magnetic field $\vec{B}(t)$.
The polarization operator $\Pi^K_{\perp,ij}$ is given by
\req{PKperp=}.

\section{Langevin equation}\label{sec:langevin}

\subsection{Langevin equation for the direction of the internal magnetic field}

In this section we consider evolution of the direction vector
$\vec{n}$, described by the tangential terms in the action,
\req{actionperp=}. We neglect fluctuations of the magnitude of the
internal magnetic field, $\vec{\magn}_\|$, the conditions when
these fluctuations can be neglected are listed in the next
section.

We decouple the quadratic in $\vec\magn^q_\perp$ component of the
action in \req{actionperp=} by introducing an auxiliary
field~$\vec{w}(t)$ with the probability distribution
\be
\mathcal{P}[\vec{w}(t)]\propto\exp\left\{\,\frac{4i}{\mls^2}\int{d}t\,dt'\,
(\Pi^K_\perp)^{-1}_{ij}(t,t')\,w_i(t)\,w_j(t')\right\},
\ee
and the correlation function
\begin{equation}
\langle{w}_i(t)\,w_j(t')\rangle=\frac{\mls^2}8\,i\Pi^K_{\perp,ij}(t,t')\,.
\end{equation}
The field $\vec{w}(t)$ plays the role of the gaussian random
Langevin force. Integration of the tangential part of the action,
\req{actionperp=}, over $\vec\magn_\perp^q$ produces a functional
$\delta$-function, whose argument determines the equation of motion:
\begin{equation}
\frac{\dot{\vec{n}}
+2\tau \magn_0 [{\vec{n}}\times\dot{{\vec{n}}}]}
{4\tau^2\magn_0^2+1}-\frac{1}{\tau\magn_0}\left(\vec{w}-\vec{n}\times[\vec{n}\times\vec{B}]\right)
=0.
\label{Langevin=}
\end{equation}
The above equation can be resolved with respect to $\dot{\vec{n}}$:
\begin{equation}
\dot{\vec{n}}=-2[\vec{n}\times(\vec{w}+\vec{B})]
-\frac{1}{\magn_0\tau}\left[\vec{n}\times[\vec{n}\times(\vec{w}+\vec{B})]\right].
\label{Langn=}
\end{equation}
This equation is the Langevin equation for the direction~$\vec{n}(t)$
of the internal magnetic field in the presence
of the external magnetic field~$\vec{B}(t)$ and the Langevin
stochastic forces $\vec{w}(t)$.

\subsection{The Fokker-Plank equation}

Next, we follow the standard procedure of derivation of the
Fokker-Plank equation for the distribution~$\mathcal{P}(\vec{n})$
of the probability for the internal magnetic field to point in
the direction~$\vec{n}$. The probability distribution satisfies the
continuity equation:
\be
\frac{\partial\mathcal{P}}{\partial{t}}+\frac{\partial J_i}{\partial
\vec{n}_i}=0,
\ee
where the probability current is defined as
\be
\begin{split}
\vec{J}=& -\left(2\vec{n}\times\vec{B}
+\frac{1}{\magn_0\tau}\vec{n}\times[\vec{n}\times\vec{B}]
\right)\mathcal{P}\\
&+\frac{1}{2}\Big{\langle}\vec{\xi} \left(\vec{\xi}\cdot\frac{\partial
\mathcal{P}}{\partial\vec{n}}\right)\Big{\rangle}
\end{split}
\label{FP1=}
\ee
and the stochastic velocity $\vec\xi$ is introduced in terms of the field $\vec{w}$
as
\be
\vec{\xi}=-2[\vec{n}\times\vec{w}]
-\frac{1}{\magn_0\tau}\left[\vec{n}\times[\vec{n}\times\vec{w}]\right].
\ee
The derivative $\partial/\partial\vec{n}$ is understood as the
differentiation with respect to local Euclidean coordinates in the
tangent space.
Performing averaging over fluctuations of $\vec{w}$ in \req{FP1=},
we obtain
\begin{equation}
\begin{split}
\frac{\partial\mathcal{P}}{\partial{t}}=&\frac{\partial}{\partial\vec{n}}
\left\{\frac{(2\magn_0\tau)[\vec{n}\times\vec{B}]
+[\vec{n}\times[\vec{n}\times\vec{B}]]}{\magn_0\tau}\,\mathcal{P}\right\}\\
&+\frac{1}{\mathcal{T}_0}
\frac{\partial^2\mathcal{P}}{\partial\vec{n}^2},
\end{split}
\end{equation}
where the time constant $\mathcal{T}_0$ is defined as
\be\label{T0=}
\mathcal{T}_0=\frac{2(\magn_0\tau)^2}{\tau{T}_{\mathrm{eff}}\mls}\,.
\end{equation}

Below we use the polar coordinates for the direction of the
internal magnetic field,
$\vec{n}=\{\sin\theta\cos\varphi,\sin\theta\sin\varphi,\cos\theta\}$.
In this case the Fokker-Plank equation can be rewritten in the
form
\be
\begin{split}
\frac{\partial\mathcal{P}}{\partial{t}}=&
\frac{1}{\sin\theta}\,\frac{\partial}{\partial\varphi}
\left[F_\varphi\mathcal{P}
+\frac{1}{\mathcal{T}_0}\,
\frac{1}{\sin\theta}\,\frac{\partial\mathcal{P}}{\partial\varphi}\right]
\\
&+\frac{1}{\sin\theta}\,\frac{\partial}{\partial\theta}
\left[\sin\theta\,F_\theta\mathcal{P}
+\frac{\sin\theta}{\mathcal{T}_0}\,
\frac{\partial\mathcal{P}}{\partial\theta}\right],\label{FPpol=}
\end{split}
\ee
where
\begin{eqnarray}
F_\varphi& = &
\frac{B_x\sin\varphi-B_y\cos\varphi}{\magn_0\tau}\nonumber
\\
&+&2\cos\theta(B_x\cos\varphi+B_y\sin\varphi)
-2\sin\theta\,B_z,\\
F_\theta & = &
2(B_x\sin\varphi-B_y\cos\varphi)\nonumber\\
&-&\frac{\cos\theta}{\magn_0\tau}
(B_x\cos\varphi+B_y\sin\varphi)
+\frac{\sin\theta}{\magn_0\tau}\,B_z.\qquad
\end{eqnarray}
It should be supplemented by the normalization condition:
\begin{equation}
\int\limits_0^{2\pi}d\varphi\int\limits_0^\pi\sin\theta\,d\theta\,
\mathcal{P}(\varphi,\theta)=1\,,
\end{equation}
which is preserved if the boundary conditions at $\theta=0,\pi$
are imposed:
\begin{equation}\label{boundarycondition=}
\lim\limits_{\theta\to{0},\pi}\sin\theta\int\limits_0^{2\pi}d\varphi\,
\frac{\partial\mathcal{P}}{\partial\theta}=0\,.
\end{equation}

Below we apply the Fokker Plank equation for calculations of the
magnetization of a particle
\be
\vec{M}=\int\frac{d\Omega_{\vec{n}}}{4\pi} \vec{n}\mathcal{P}(\vec{n})
\label{Mdef=}
\ee

\section{Applicability of the approach}\label{sec:applicability}

In this section we discuss the conditions of validity of the
stochastic LLG equation, see \req{FPpol=}, for the model of ferromagnetic
metallic particle connected to leads at finite bias. We briefly
listed these conditions in the Introduction. Here we present their
more detailed quantitative analysis.

\subsection{Fluctuations of the radial component of the internal magnetic field}

We represented the classical component of the internal magnetic
field $\vec{\magn}^{cl}$ in terms of a slowly varying direction
$\vec{n}(t)$ and fast oscillations $\magn_\|^{cl}$ of its magnitude
around the average value $h_0$. Now, we evaluate the amplitude of
oscillations of the radial component $\magn_\|^{cl}$ of the field,
using the radial term in the action, see \reqs{actiontot=} and
\rref{actionrad=}.

The typical frequencies for time evolution of small fluctuations
of the internal magnetic field in the radial direction are of order of
\be
\label{magnflucrate=}
\omega\sim \frac{\mls+E''/2}{\mls} \,\frac{1}{\tau}
\ee
as one can conclude from the explicit form of the propagator
$\mathcal{D}^R_\|(\omega)$, \req{DparrR=}, of these fluctuations.
This scale has the meaning of the inverse $RC$-time in the spin
channel. Deep in the ferromagnetic state (i.~e., far from the Stoner
critical point $E''+2\mls=0$) we estimate $\mls+E''/2\sim\mls$
(which is equivalent to $E''\sim\magn_0/S_0$), so this spin-channel
$RC$-time is of the same order as the escape time~$\tau$.
This estimate for the frequency
range is consistent with the simple picture, which describes the
evolution of the internal magnetic field of the grain as a response
to a changing value of the total spin of the particle due to random
processes of electron exchange between the dot and the leads. The
electron exchange happens with the characteristic rate $1/\tau$.

The correlation function $\langle
\magn_\|^{cl}(t)\magn_\|^{cl}(t')\rangle$ can be evaluated by
performing the Gaussian integration with the quadratic action in
$\magn_\|^{cl}$ and $\magn_\|^{q}$. Using \req{DparrK=}, we obtain
the equal-time correlation function
\begin{equation}\begin{split}
&\langle(\magn_\|^{cl})^2\rangle
=\frac{i}2\,\int\frac{d\omega}{2\pi}\,\mathcal{D}^K_\|(\omega)\\
&=\frac{(E'')^2}{32\tau\mls}\int\frac{d\omega}{2\pi}\,
\frac{2\pi\mathcal{R}(\omega)}
{\omega^2+[1+E''/(2\mls)]^2/\tau^2}.
\end{split}\end{equation}
This equation gives the value of fluctuations of the radial
component of the internal magnetic field of the particle. These
fluctuations survive even in the limit $T=0$ and $V=0$, when
$\mathcal{R}(\omega)=2|\omega|/\pi\tau$. We have the following
estimate
\begin{equation}
\langle(\magn_\|^{cl})^2\rangle
=\frac{(E'')^2}{16\pi\tau\mls}
\ln\frac{E_{\rm T}\tau}{1+E''/(2\mls)}\,,
\label{hradfl=}
\end{equation}
the upper cutoff $E_{\rm T}$ is the Thouless energy, $E_{\rm T}= v_F/L$
for a ballistic dot with diameter $L$ and electron Fermi velocity $v_F$.

The separation of the internal magnetic field into the radial and
tangential components is justified, provided that the fluctuations
$\sqrt{\langle(\magn_\|^{cl})^2\rangle}$ of the radial component are
much smaller than the average value of the field $\magn_0$, i.e.
$\langle(\magn_\|^{cl})^2\rangle\ll \magn_0^2$. Using the estimate
\req{hradfl=}, we obtain the necessary requirement for the
applicability of equations for the slow evolution of the vector of
the internal magnetic field of a particle:
\begin{equation}
S_0\gg \sqrt{\frac{1}{\tau\mls}\ln(E_{\rm T}\tau)},
\label{condA>}
\end{equation}
where $S_0$ is the spin of a particle in equilibrium and we again
used the estimate ${E}''\sim\magn_0/S_0$.
Condition of \req{condA>} requires that the system is not close to
the Stoner instability.

\subsection{Applicability of the gaussian approximation}\label{sec:gaussian}

Let us discuss the applicability of the gaussian approximation
for the action in $\magn_\|^{cl}$ and $\vec\magn^q$. The
coefficients in front of terms
$\vec\magn^q(t)\,\magn_\|(t_1)\ldots\magn_\|^{cl}(t_n)$
are obtained by taking the $n$th variational derivative of
$\delta\vec{g}^K(t,t)+\delta\vec{g}^Z(t,t)$, or, equivalently, by
iterating the Usadel equation $n$~times. Since the typical frequencies
of $\magn_\|$ are $\omega\sim{1}/\tau$, the left-hand side of
the equation is $\sim\delta^{(n+1)}\vec{g}^K/\tau$, while the
right-hand side is $\magn_\|^{cl}\delta^{(n)}\vec{g}^K$. Since the
only time scale here is~$\tau$, all the coefficients of the expansion of
the action in $\magn_\|^{cl}(\omega)$ at
$\omega\sim{1}/\tau$ are of the same order:
\begin{eqnarray}
&&S_{n+1}\sim\frac{\tau^{n-1}}{E''}
\int\frac{d\omega_1\ldots{d}\omega_n}{(2\pi)^n}
\times\nonumber\\
&&\qquad{}\times\magn_\|^{cl}(\omega_1)\ldots\magn_\|^{cl}(\omega_n)\,
\magn^q(-\omega_1-\ldots-\omega_n).\qquad
\end{eqnarray}
At the same time, the typical value of $\magn_\|^{cl}(\omega\sim{1}/\tau)$,
as determined by the gaussian part of the action, was estimated in the
previous subsection to be of the order of
$\sqrt{\tau\mathcal{D}^K_\|(\omega\sim\tau)}\sim\sqrt{\tau\mls}\ll{1}$,
so the higher-order terms are indeed not important.

For the quantum component of the field
the quadratic and quartic terms in the action are estimated as
\begin{eqnarray}
&&S_{n+1}\sim\frac{T_\mathrm{eff}\tau^n}\mls
\int\frac{d\omega_1\ldots{d}\omega_n}{(2\pi)^n}
\times\nonumber\\
&&\qquad{}\times\magn^q(\omega_1)\ldots\magn^q(\omega_n)\,
\magn^q(-\omega_1-\ldots-\omega_n).\qquad
\end{eqnarray}
If $T_\mathrm{eff}\gg{1}/\tau$, then the typical frequency scale is $\omega\sim{1}/\tau$,
so the quadratic term gives $\magn^q(\omega\sim{1}/\tau)\sim\sqrt{\mls/T_\mathrm{eff}}$,
and $S_n\sim(\mls/T_\mathrm{eff})^{n/2-1}\sim[\tau\mls/(\tau{T}_\mathrm{eff})]^{n/2-1}$.
If $T_\mathrm{eff}\ll{1}/\tau$,
at the typical scale $\omega\sim{T}_\mathrm{eff}$ we obtain
$\magn^q(\omega\sim{T}_\mathrm{eff})\sim\sqrt{\mls/(T_\mathrm{eff}^2\tau)}$,
so again $S_n\sim(\tau\mls)^{n/2-1}\ll{1}$ for $n>2$.

Physically, the parameter $1/(\tau\mls)=N_{ch}$ (or $T_\mathrm{eff}/\mls$,
if it is larger) can be identified with the number of the independent sources
of the noise acting on the magnetization field. Thus, the smallness of the
non-gaussian part of the action is nothing but the manifestation of the
central limit theorem.

\subsection{Applicability of the Fokker-Plank equation}
From the above analysis we found that evolution of the direction of
the internal magnetic field in time is described by a characteristic
time $\mathcal{T}_0$, introduced in \req{T0=}. From the analysis of
the fluctuations of the magnitude of the internal magnetic field,
see \req{magnflucrate=}, we obtain the following condition when the
separation into slow and fast variables is legitimate. The criterium
can be formulated as $\mathcal{T}_0\gg \tau$, which can be presented
as
\be
\frac{T_{\rm eff}}{\mls}\ll
\left(\frac{\magn_0}{\mls}\right)^2 = S_0^2.
\label{slowfast=}
\ee

\section{Magnetic susceptibility of metallic particles out of equilibrium}
\label{sec:susceptibility}
The LLG equation derived in this paper for a ferromagnetic
particle with finite bias between the leads can be applied to a
number of experimental setups. Moreover, the derivation of the
equation can be generalized to spin-anisotropic contacts with
leads or Hamiltonian of electron states in the particle. In this
paper we apply the stochastic equation for spin distribution
function to the analysis of the magnetic susceptibility at finite
frequency. The susceptibility is the basic characteristic of
magnetic systems, it can often be measured directly, and determines
other measurable quantities.

Below, we calculate the susceptibility of an ensemble of particles
placed in constant magnetic field of an arbitrary strength
and oscillating weak magnetic field, see Fig.~1. We consider the
oscillating magnetic field with its components in directions
parallel and perpendicular to the constant magnetic field.

\subsection{Solution at zero noise power}

At $T_{\mathrm{eff}}=0$ when $\vec{w}(t)=0$, and at fixed direction
of the field, $\vec{B}(t)=\vec{e}_zB(t)$, equation of motion~(\ref{Langn=})
is easily integrated for an arbitrary time dependence~$B(t)$:
\begin{eqnarray}
\varphi &=&
\varphi_0+\int\limits_0^t{2}B(t')\,dt',\\
\tan\frac\theta{2}  & = & \tan\frac{\theta_0}{2}
\exp\left[-\int\limits_0^t\frac{B(t')}{\magn_0\tau}\,dt'\right].
\end{eqnarray}
Here the direction of magnetic field corresponds to $\theta=0$.

\subsection{Constant magnetic field}

At finite $T_{\mathrm{eff}}$ in constant magnetic field $B_0$ the
Fokker-Plank equation has a simple solution
\begin{equation}
\mathcal{P}_0(\theta)=\frac{b}{\sinh{b}}\frac{e^{b\cos\theta}}{4\pi}\,,
\label{P0=}
\end{equation}
where the strength of constant magnetic field is
written in terms of the dimensionless parameter
\begin{equation}
b\equiv\frac{(2\magn_0\tau)B_0}{\tau\mls{T}_{\mathrm{eff}}}\,.
\label{b=}
\end{equation}
Substituting this probability function to \req{Mdef=}, we obtain
the classical Langevin expression for
the magnetization of a particle in a magnetic field
\be
M_z=
\coth{b}-\frac{1}b\,, \quad M_x=M_y=0.
\ee
This expression for the magnetization coincides with the
magnetization of a metallic particle in thermal equilibrium,
provided that the temperature is replaced by the effective
temperature $T_\mathrm{eff}$ defined by \req{Teff=}.

The differential dc susceptibility is equal to
\be
\label{chiparrdc=}
\chi_\|^{\rm dc}=\frac{dM_z(b)}{db}=\frac{1}{b^2}-\frac{1}{\sinh^2
b}.
\ee

\subsection{Longitudinal susceptibility}

We now consider the response of the magnetization to weak
oscillations $\tilde B_z(t)$ of the external magnetic field with frequency $\omega$
in direction parallel to the fixed magnetic field $B_0$. We write
the oscillatory component of the field in terms of the dimensionless
field strength:
\begin{equation}
b_\|\,{e}^{-i\omega{t}}+b_\|^*{e}^{i\omega{t}}=
\frac{2\magn_0\tilde B_z(t)}{\mls{T}_{\mathrm{eff}}}.
\end{equation}
The linear correction to the probability distribution can be cast in
the form
\begin{equation}
\mathcal{P}(\theta,t)=\left[1+b_\|u_\|(\theta)\,e^{-i\omega{t}}
+b_\|^*u_\|^*(\theta)\,e^{i\omega{t}}\right]\mathcal{P}_0(\theta)\,,
\label{Pparr=}
\end{equation}
with $\mathcal{P}_0(\theta)$ defined by \req{P0=}. The magnetic ac
susceptibility can be evaluated from \req{Pparr=} as
\begin{equation}
\chi_\|(\omega,b)=2\pi \int\limits_{0}^\pi u_\|(\theta) \mathcal{P}_0(\theta) \cos\theta \sin\theta d\theta\,.
\label{chi||=}
\end{equation}

The equation for $u_\|(\theta)$ is obtained from \req{FPpol=} with $B_z=B_0+\tilde
B_z(t)$:
\begin{equation}
\frac{\partial^2u_\|}{\partial\theta^2}
+\frac{\cos\theta-b\sin^2\theta}{\sin\theta}\,\frac{\partial{u}_\|}{\partial\theta}
+i\Omega{u}_\|=b\sin^2\theta-2\cos\theta\,,
\label{longitudinaldiffur=}
\end{equation}
where we introduced the dimensionless frequency
\be\label{Omega=}
\Omega=\omega\mathcal{T}_0\,,
\ee
and the time constant $\mathcal{T}_0$ is defined in \req{T0=}.

Note
the symmetry of Eq.~(\ref{longitudinaldiffur=}) with respect to the
simultaneous change $b\to-b$ and $\theta\to\pi-\theta$.
Also, the normalization condition for the probability function
requires that
\begin{equation}
\int\limits_{0}^\pi u_\|(\theta) \mathcal{P}_0(\theta)  \sin\theta d\theta=0\,.
\label{normparr=}
\end{equation}
The latter holds if the boundary conditions \req{boundarycondition=}
are satisfied, which in the case of axial symmetry can be written as
\be
\label{boundaryconditionparr=}
\lim\limits_{\theta\to{0},\pi}\left\{\sin\theta
\frac{\partial u_\|(\theta)}{\partial\theta}\right\}=0\,.
\ee

\begin{figure}
\epsfxsize=0.4\textwidth
\centerline{\epsfbox{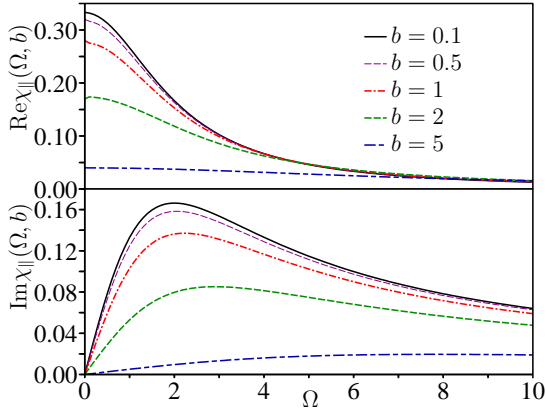}}
\caption{(Color online). Plot of the real and imaginary parts of the susceptibility $\chi_\|(\Omega, b)$
as a function of the dimensionless frequency $\Omega=\omega{\cal T}_0$.
The oscillatory field at frequency $\omega$ is parallel to the constant magnetic field with strength $b$.
The real part of the susceptibility decreases monotonically from its dc value, \req{chiparrdc=},
as frequency increases, while the imaginary part increases linearly at small $\Omega\ll 1$, see \req{chiparrlf=},
and decreases at higher frequencies.}
\label{fig:2}
\end{figure}

\begin{figure}
\epsfxsize=0.4\textwidth
\centerline{\epsfbox{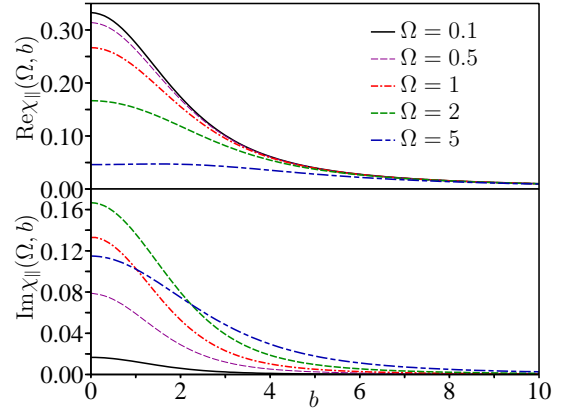}}
\caption{(Color online). Plot of the real and imaginary parts of the $ac$ susceptibility $\chi_\|(\Omega, b)$
at several values of the dimensionless frequency $\Omega$ of the oscillating
magnetic field along the constant magnetic field with strength
$b$. In general, magnetic field suppresses both real and imaginary parts of the susceptibility.
}
\label{fig:3}
\end{figure}

The differential equation \rref{longitudinaldiffur=} with the
boundary condition \req{boundaryconditionparr=} can be solved
numerically and then the susceptibility is evaluated according to
\req{chi||=}. The result is shown in Figs.~\ref{fig:2} and
\ref{fig:3}, where the susceptibility is shown as a function of
frequency $\omega$ or magnetic field $b$, respectively.
We also consider various asymptotes for the ac susceptibility,
obtained from the solution of \req{longitudinaldiffur=}.

At zero constant magnetic field, $b=0$,
we find the exact solution of \req{longitudinaldiffur=} explicitly:
\begin{equation}
u_\|(\theta)=\frac{\cos\theta}{1-i\Omega/2}\,.
\label{uparrzb=}
\end{equation}
This solution allows us to calculate the ac susceptibility in the
form
\be
\label{chiparrzb=}
\chi_\|(\Omega,b=0)=\frac{1}{3}\frac{1}{1-i\Omega/2}\,.
\ee

For $b\gg{1}$ only $\cos\theta\sim{1}/b$ matter, and we can find a
specific solution of the inhomogeneous equation:
\begin{equation}
u_\|(\theta)=\frac{1-b(1-\cos\theta)}{b-i\Omega/2}\,,\quad
b\gg{1}.
\end{equation}
The requirement of regularity at the opposite end can be replaced by the
probability normalization condition, \req{normparr=},
which is satisfied by this solution. Substituting this solution to
\req{chi||=}, we obtain the strong field asymptote for the ac
susceptibility
\be
\label{chiparrhb=}
\chi_\|(\Omega,b)=\frac{1}{b(b-i\Omega/2)}\,.
\ee

For $\Omega\gg{1}$ and $\Omega\gg{b}$,  we can neglect the derivatives in  \req{longitudinaldiffur=}
and find the solution in the form
\begin{equation}
u_\|(\theta)\approx\frac{b\sin^2\theta-2\cos\theta}{i\Omega}\,,\quad
\end{equation}
This solution $u_\|(\theta)$ also satisfies \req{normparr=}.
For the susceptibility, \req{chi||=}, we obtain
\be
\chi_\|(\Omega\to\infty,b)=
\frac{2i}{\Omega}\left(\frac{\coth{b}}{b}-\frac{1}{b^2}\right).
\label{chiparrhf=}
\ee

\begin{figure}
\epsfxsize=0.4\textwidth
\centerline{\epsfbox{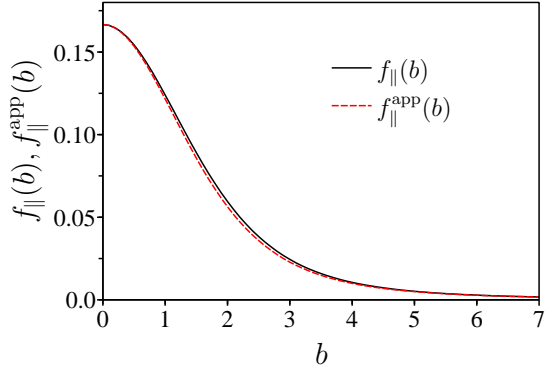}}
\caption{(Color online). Dependence on magnetic field $b$ of the
slope of the imaginary part of the linear in frequency  susceptibility
$\chi_\|(\Omega, b)$ at low frequencies $\Omega\ll 1$, calculated
according to \req{chiparrlf=}. For comparison, we also plot
function $f_\|^{\rm app}(b)$, see \req{fparr=}.
}
\label{fig:4}
\end{figure}

Finally, the low frequency limit can be also analyzed
analytically. The real part of the susceptibility coincides with
the differential susceptibility in dc magnetic field,
\req{chiparrdc=}, for the imaginary part to the first order in
frequency we obtain, see Appendix,
\be
{\rm Im}\chi_\|(\Omega,b)=\Omega f_\|(b).
\label{chiparrlf=}
\ee
The function $f_\|(b)$ has a complicated analytical form and is not
presented here, but its plot is shown in Fig.~\ref{fig:4}.

In all considered four limiting cases, the asymptotic approximations
hold regardless the order in which the limits are taken. Indeed,
the asymptote of the expression for the susceptibility
in the zero field, \req{chiparrzb=}, has the asymptote at
$\Omega\to\infty$ consistent with \req{chiparrhf=} at
$b=0$. Similarly, the high frequency limit of \req{chiparrhb=}
coincides with the limit $b\to\infty$ of \req{chiparrhf=}. Both
limits of weak and strong magnetic field of the imaginary part of the susceptibility
at low frequencies, \req{chiparrlf=}, coincide with the imaginary
part of $\chi_\|(\Omega,b)$, calculated from \req{chiparrzb=} and
\req{chiparrhf=}, respectively.

In
general, we make a conjecture that the ac susceptibility is given
by the following expression:
\be
\chi_\|(\Omega,b)=\sum_n\frac{\chi^\|_n(b)}{1-i\Omega/\Gamma^\|_n(b)},
\ee
where functions $\chi^\|_n(b)$ and $\Gamma^\|_n(b)$ are real and
describe the degeneracy points of the homogeneous differential
equation \req{longitudinaldiffur=} with real $i\Omega$. This
expansion is related to the expansion of time-dependent Fokker-Plank
equations in the spherical harmonics, analyzed in
Ref.~\onlinecite{Brown1963}. In particular,
$\chi^\|_{n>1}(b\to 0)=O(b)$ and $\Gamma^\|_n(b\to 0)=n(n+1)+O(b)$.

For practical purposes, we found from a numerical analysis that even
the single-pole approximation,
\be
\chi_\|^{\rm app}(\omega,b) = \left(\frac{1}{b^2}-\frac{1}{\sinh^2b}\right)
\left[1-i\omega\mathcal{T}_\|(b)\,\right]^{-1},
\label{chiparrcnj=}
\ee
gives a very good estimate of the susceptibility for all values of
$\omega$ and $b$. The analysis shows
that the susceptibility, \req{chi||=}, obtained from a numerical solution
of \req{longitudinaldiffur=}, is within a few per cent of the estimate given by
\req{chiparrcnj=}.
The characteristic time constant, $\mathcal{T}_\|(b)$,
as a function of magnetic field~$b$ is chosen from the
high frequency asymptote \req{chiparrhf=}:
\be
\mathcal{T}_\|(b)=\frac{\mathcal{T}_0}{2\sinh{b}}\,\frac{\sinh^2b-b^2}{b\cosh{b}-\sinh{b}}\,.
\label{tauparr=}
\ee
To evaluate the accuracy of the above approximation,
\req{chiparrcnj=}, we consider the opposite limit of low frequencies, $\Omega\ll 1$,
and compare the exact result for the
imaginary part of the susceptibility, \req{chiparrlf=}, with
\be
\begin{split}\label{fparr=}
&{\rm Im}\chi_\|^{\rm app}(\omega,b)  =\omega \mathcal{T}_0 f_\|^{\rm
app}(b),
\\
&f_\|^{\rm app}(b) =
\frac{1}{2b^2\sinh^3{b}}\,\frac{(\sinh^2b-b^2)^2}{b\cosh{b}-\sinh{b}}\,.
\end{split}\ee
For visual comparison of functions $f_\|(b)$ and $f_\|^{\rm app}(b)$, we
plot both functions in Fig.~\ref{fig:4}, where these curves are nearly indistinguishable.
The difference between these two
curves vanishes at $b\to 0$ and $b\to\infty$, and has a maximal
difference at $b\approx 2$, which constitutes only tiny fraction
of $f_\|(b)$.

\subsection{Transverse susceptibility}

Next, we consider the response of the magnetization to weak
oscillations $\tilde{\vec{B}}_\perp(t)$ of the external magnetic field with frequency $\omega$
in direction perpendicular to the fixed magnetic field $B_0$. We write
the oscillatory component of the field in the form:
\begin{equation}
\tilde{\vec{B}}_\perp(t)=\frac{\mls{T}_{\mathrm{eff}}}{2\magn_0}\left[
b_\perp(\vec{e}_x+i\vec{e}_y)\,{e}^{-i\omega{t}}
+b_\perp^*(\vec{e}_x-i\vec{e}_y){e}^{i\omega{t}}\right].
\label{Bperp=}
\end{equation}
This field represents a circular polarization of an ac magnetic field
in the $(x,y)$ plane, perpendicular to the fixed magnetic field in the
$z$-direction: $\vec{B}=\{B_\perp\cos\omega t; B_\perp\sin\omega t; B_0\}$.
We look for the linear correction to the probability distribution
in the form
\begin{equation}\begin{split}
\mathcal{P}&(\varphi,\theta,t)= \mathcal{P}_0(\theta)\\
&\times\left[1+b_\perp{u}_\perp(\theta)\,e^{i\varphi-i\omega{t}}
+b_\perp^*{u}_\perp^*(\theta)\,e^{-i\varphi+i\omega{t}}\right].
\end{split}\end{equation}
The equation for $u_\perp(\theta)$ is obtained from
the  Fokker-Plank equation  \req{FPpol=}, linearized in the parameter~$b_\perp$:
\begin{equation}\begin{split}
\frac{\partial^2u_\perp}{\partial\theta^2}
+&\frac{\cos\theta-b\sin^2\theta}{\sin\theta}\,\frac{\partial{u}_\perp}{\partial\theta}+
\left(i\Omega_\perp-\frac{1}{\sin^2\theta}\right){u}_\perp
\\
&=-\sin\theta\left(2+2i\magn_0\tau{b}+b\cos\theta\right).
\label{transversediffur=}
\end{split}\end{equation}
Here the dimensionless frequency is a difference between the drive
frequency $\omega$ and the precession frequency in external field
$B_0$:
\be
\Omega_\perp=(\omega-2B_0)\,\mathcal{T}_0=\Omega-2 (\magn_0\tau) b,
\ee
where $\mathcal{T}_0$ is defined in \req{T0=} and the right equality is written in terms of
dimensionless variables $\Omega$, \req{Omega=}, and $b$, \req{b=}.
Equation \rref{transversediffur=} is symmetric with respect to the simultaneous change
$\theta\to\pi-\theta$, $b\to-b$, $i\to-i$, $\Omega_\perp\to-\Omega_\perp$ (``parity'').
The function~$\mathcal{P}(\varphi,\theta,t)$ is single-valued at the poles
$\theta=0$ and $\theta=\pi$, only if
\be\label{boundaryconditionperp=}
u_\perp(\theta=0)=0, \quad
u_\perp(\theta=\pi)=0.
\ee
The latter equations establish the boundary
conditions for the differential equation~\rref{transversediffur=}.
We also note that the normalization condition is satisfied for any
function $u_\perp(\theta)$.

We define the susceptibility in response to the ac magnetic field,
\req{Bperp=}, as
\be
\chi_\perp(\Omega,b)=2\pi \int\limits_{0}^\pi u_\perp(\theta) \mathcal{P}_0(\theta)
\sin^2\theta d\theta\,.
\label{chiperp=}
\ee
This expression for the susceptibility can be used to calculate the
magnetization of a particle
\be
\vec{M}(t)=\int
\vec{n}(\varphi,\theta)\mathcal{P}(\varphi,\theta,t)\sin\theta\,d\theta
d\varphi
\ee
to the lowest order in the ac magnetic field. In particular,
\be
M_x(t)=\Re(\chi_\perp{b}_\perp{e}^{-i\omega{t}}),\quad
M_y(t)=\Im(\chi_\perp{b}_\perp{e}^{-i\omega{t}}).
\ee

\begin{figure}{t}
\epsfxsize=0.45\textwidth
\centerline{\epsfbox{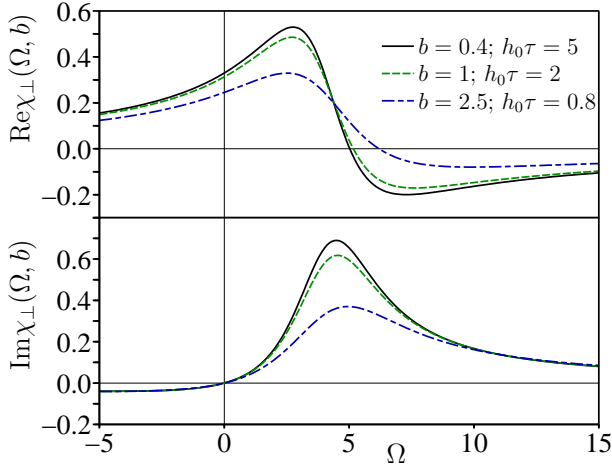}}
\caption{(Color online). Plot of the real and imaginary parts of the transverse susceptibility $\chi_\perp(\Omega, b)$
as a function of the dimensionless frequency $\Omega$.
Negative frequency corresponds to the opposite sense of the
circular polarization of the ac magnetic field in a plane,
perpendicular
to the constant magnetic field with strength $b$. The parameters
of the three shown curves are chosen so that $\magn_0\tau b =2$.
}
\label{fig:5}
\end{figure}

\begin{figure}
\epsfxsize=0.45\textwidth
\centerline{\epsfbox{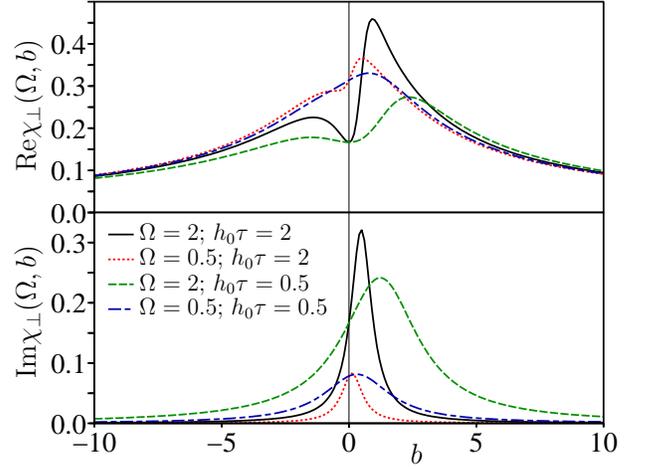}}
\caption{(Color online). Plot of the real and imaginary parts of the transverse susceptibility $\chi_\perp(\Omega, b)$
as a function of the strength $b$ of a constant magnetic field,
shown for two values of frequency $\Omega$ and two values of the ``damping factor''
$\magn_0\tau$.
Negative values of $b$ corresponds to the opposite sense of the
circular polarization of the ac magnetic field in a plane,
perpendicular
to the constant magnetic field. The real part of the susceptibility exhibits a
strong non-monotonic behavior at weak magnetic fields.}
\label{fig:6}
\end{figure}

Solving numerically the differential equation \rref{transversediffur=} with the
corresponding boundary conditions, \req{boundaryconditionperp=}, we
obtain the transverse susceptibility, \req{chiperp=},
shown in Figs.~\ref{fig:5} and \ref{fig:6}.
Below we analyze several limiting cases.

In zero fixed magnetic field, $b=0$, we have the  exact solution
of \req{transversediffur=}:
\be
u_\perp(\theta)=\frac{\sin\theta}{1-i\Omega_\perp/2}\,.
\ee
This solution corresponds to the solution in the longitudinal case,
rotated by 90$^\circ$, cf. \req{uparrzb=}.

At $\omega=0$ and $\Omega_\perp=-2 b\magn_0\tau$, the solution of
\req{transversediffur=} has a simple form
$u_\|(\theta)=\sin\theta$ and corresponds to a tilt of the external
field. The susceptibility due to such tilt is
\be
\chi_\perp(\Omega=0,b)=\frac{2}{b^2}\left(b\coth b-1\right).
\ee

In strong fixed magnetic field, $b\gg{1}$, we need to consider
small angles $\theta\sim{1}/\sqrt{b}$, therefore, we can approximate
$\cos\theta\approx 1$ in \req{transversediffur=} and obtain:
\be
u_\perp(\theta)=\frac{b+2+2i\magn_0\tau{b}}{b+2-i\Omega_\perp}\,\sin\theta\,.
\ee
The susceptibility in the limit $b\gg 1$ is given by
\be
\chi_\perp(\Omega,b) =\frac{1+2i\magn_0\tau}{b(b(1+2i\magn_0\tau)-i\Omega)}\,.
\ee

At $\Omega_\perp\gg{1},b$ we can disregard the terms in
\req{transversediffur=} with derivatives. Moreover, the
contribution to the susceptibility, \req{chiperp=}, from the vicinity of
$\theta=0$ and $\theta=\pi$ is suppressed as $\sin^2\theta$. This
observation allows us to write the solution in the form
\be
u_\perp(\theta)=\sin\theta\,\frac{2+2i\magn_0\tau{b}+b\cos\theta}{-i\Omega_\perp},
\ee
Consequently, we obtain the following high frequency, $\Omega_\perp\gg
1$, asymptote for the susceptibility:
\be
\chi_\perp(\Omega,b)=\frac{i}{\Omega-2 \magn_0\tau b}\left[\frac{b\coth{b}-1}{b^2}\,(2i\magn_0\tau{b}-1)+1\right]
\,.
\ee

We can use the approximate expression for the susceptibility in
response to the transverse oscillating magnetic field
\be\begin{split}
\chi_\perp^{\rm app}(\omega) =&\frac{b\coth{b}-1}{b^2}
\left[1+2iB_0\mathcal{T}_\perp(b)\,\right]\\
&\times \left[1+i(2B_0-\omega)\mathcal{T}_\perp(b)\right]^{-1}\nonumber
\end{split}\ee
The corresponding characteristic time constant   $\mathcal{T}_\perp(b)$
can be found for any~$b$ from the asymptotic
behavior of $\chi_\perp(\Omega,b)$ at $\Omega_\perp\gg{1,b}$:
\begin{equation}
\mathcal{T}_\perp(b)=\mathcal{T}_0\, \frac{b\coth{
b}-1}{b^2+1-b\coth{b}}\,.
\end{equation}

\section{Conclusions}

We have studied the slow dynamics of magnetization in a small metallic
particle (quantum dot), where the ferromagnetism has arisen as a consequence
of Stoner instability.
The particle is connected to non-magnetic electron reservoirs. A
finite bias is applied between the reservoirs, thus bringing the whole
electron system away from equilibrium. The exchange of electrons
between the reservoirs and the particle
results in the Gilbert damping\cite{Gilbert55}
of the magnetization dynamics and  in
a temperature- and bias-driven Brownian motion of the direction of the particle
magnetization.
Analysis of magnetization dynamics and transport properties
of ferromagnetic nanoparticles is commonly
performed\cite{Palacios1998,Usadel2006,Foros2007,Denisov2007,Kamenev2008} within
the stochastic Landau-Lifshitz-Gilbert (LLG) equation~\cite{LL35,Gilbert55}, which is an
analogue of the Langevin equation written for a unit
three-dimensional vector.

We derived the stochastic LLG
 equation from a microscopic starting point and
established conditions under which the description of
the magnetization of a ferromagnetic metallic particle by this equation is applicable.
We concluded that the applicability of the LLG equation for a ferromagnetic particle
is set by three independent criteria.
(1) The contact resistance should be low compared to the resistance
quantum, which is equivalent to $N_{ch}\gg{1}$. Otherwise the noise cannot be considered
gaussian. Each channel can be viewed as an independent source of noise and only
the contribution of many channels results in the gaussian noise by virtue
of the central limit theorem for $N_{ch}\gg{1}$. (2)
The system should not be too close to the Stoner instability: the mean-field
value of the total spin $S_0^2\gg N_{ch}$. Otherwise,
the fluctuations of the absolute value of the magnetization become of
the order of the magnetization itself. (3) $S_0^2\gg{T}_\mathrm{eff}/\mls$,
where $T_\mathrm{eff}\simeq {\rm max}\{T,|eV|\}$ is the effective temperature of the system, which is
the energy scale of the electronic distribution function. Otherwise, the separation
into
slow (the direction of the magnetization) and fast (the electron
dynamics and the magnitude of the
magnetization) degrees of freedom
is not possible.

Under the above conditions, the dynamics of the magnetization  is
described in terms of the stochastic LLG equation with the power of
Langevin forces determined by the effective temperature of the system. The effective temperature
is the characteristic energy scale of the electronic distribution function in
the particle determined by a combination of the temperature and the bias
voltage. In fact, for a considered
here system with non-magnetic contacts between non-magnetic
reservoirs and a ferromagnetic particle the power of the Langevin forces
is proportional to the low-frequency
noise of total charge current through the particle.
We further reduced the stochastic LLG equation
to the Fokker-Planck equation for a unit vector, corresponding to the
direction of the magnetization of the particle. The Fokker-Plank
equation can be used to describe time evolution of the distribution
of the direction of magnetization in the presence of time-dependent
magnetic fields and voltage bias.

As an example of application of the Fokker-Plank equation for the magnetization, we have
calculated the frequency-dependent magnetic susceptibility of the particle
in a constant external magnetic field (i.~e., linear response of the
magnetization to a small periodic modulation of the field, relevant for
ferromagnetic resonance measurements). We have not been able to obtain
an explicit analytical expression for the susceptibility at arbitrary
value of the applied external field and frequency; however, analysis of
different limiting cases has lead us to a simple analytical expression
which gives a good agreement with the numerical solution of the Fokker-Planck
equation.

\section*{Acknowledgements} We acknowledge discussions with I. L.
Aleiner, G. Catelani, A. Kamenev and E. Tosatti.
M.G.V. is grateful to the International Centre for Theoretical Physics
(Trieste, Italy) for hospitality.

\appendix
\section{Longitudinal susceptibility at low frequencies }

We find the linear in frequency $\Omega\ll 1$ correction to the dc
susceptibility. For this purpose, we look for a solution to
\req{longitudinaldiffur=} in the form
\be
u_\|(\theta)=u_\|^{(0)}(\theta)+u_\|^{(1)}(\theta),
\ee
where $u_\|^{(0)}(\theta)$ is the solution of \req{longitudinaldiffur=}
at $\Omega=0$ and $u_\|^{(1)}(\theta)\propto \Omega$. We choose
\be
u_\|^{(0)}(\theta)=\frac{1}{b}-\coth b+\cos\theta,
\ee
since this form of $u_\|^{(0)}(\theta)$ preserves the normalization condition
\rref{normparr=}. This function can be found directly as a solution of
\req{longitudinaldiffur=} with $\Omega=0$ or as a variational derivative
of function $\mathcal{P}_0(\theta)$, defined in \req{P0=}, with respect
to~$b$.

The linear in $\Omega$ correction $u_\|^{(1)}(\theta)$ is the
solution to the differential equation
\be
\frac{\partial^2 u_\|^{(1)}(\theta)}{\partial \theta^2}
+\frac{\cos\theta-b\sin^2\theta}{\sin\theta}
\frac{\partial u_\|^{(1)}(\theta)}{\partial \theta}
=-i\Omega u_\|^{(0)}(\theta).
\ee
From this equation, we can easily find
\be
\frac{\partial u_\|^{(1)}(\theta)}{\partial \theta}=
-\frac{i\Omega}{b\sin\theta}\left[\coth b-\cos\theta- \frac{e^{-b\cos\theta}}{\sinh
b}\right].
\label{u1deriv}
\ee
We notice that the solution to the latter equation will
automatically satisfy the boundary conditions, given by
\req{boundaryconditionparr=}.
Integrating \req{u1deriv} once again, we obtain the following
expression for function $u_\|^{(1)}(\theta)$:
\be
\begin{split}
u_\|^{(1)}(\theta)& =C(b)
\\
- & \frac{i\Omega}{b}\int_0^\theta
\left[\coth b-\cos\theta'- \frac{e^{-b\cos\theta'}}{\sinh
b}\right]\frac{d\theta'}{\sin\theta'}.
\end{split}
\ee
Here the integration constant $C(b)$ has to be chosen to satisfy
the normalization condition, \req{normparr=}, which results in
complicated expression for the final form of the function
$u_\|^{(1)}(\theta)$.

To obtain function $f_\|(b)$, introduced in  \req{chiparrlf=}, we
have to perform the final integration
\be
f_\|(b)=\frac{2\pi}{\Omega}\int_0^\pi u_\|^{(1)}(\theta)
\mathcal{P}_0(\theta)\sin\theta\cos\theta d\theta.
\ee
The result of integration is shown in Fig.~\ref{fig:4}.

\end{document}